\documentclass[oldversion]{aa}
\usepackage{graphicx}
\usepackage{txfonts}
\begin{document}
\title{Two $\sim$35\,day clocks in Her X-1: evidence for neutron star free precession}

%   \subtitle{I. Overviewing the $\kappa$-mechanism}

\author{R.~Staubert\inst{1}, 
D.~Klochkov\inst{1},
K.~Postnov\inst{2}, N.~Shakura\inst{2}, 
J.~Wilms\inst{3}, R.E.~Rothschild\inst{4}}

\offprints{staubert@astro.uni-tuebingen.de}

\institute{
	Institut f\"ur Astronomie und Astrophysik, Universit\"at T\"ubingen,
	Sand 1, D-72076 T\"ubingen, Germany
\and
	Sternberg Astronomical Institute, 13 Universitetskii pr., 119992 Moscow, Russia
\and
        Dr.\ Remeis-Sternwarte, Astronomisches Institut der
	Universit\"at Erlangen-N\"urnberg, Sternwartstr. 7, 96049 Bamberg, Germany
\and
        Center for Astrophysics and Space Sciences, University of
        California at San Diego, La Jolla, CA 92093-0424, USA
}

%   \email{staubert@astro.uni-tuebingen.de}

\date{Received 04/08/2008; accepted 13/11/2008}
\authorrunning{Staubert et. al.}
\titlerunning{Two $\sim$35\,day clocks  in Her X-1}

% \abstract{}{}{}{}{} 
% 5 {} token are mandatory

  \abstract
   {We present evidence for the existence of two $\sim$35\,day clocks in the 
Her~X-1/HZ~Her binary system. $\sim$35\,day modulations are observed
1) in the \textsl{Turn-On} cycles with two on- and two off-states, 
and 2) in the changing  \textit{shape of the pulse profiles} which 
re-appears regularly. The two ways of counting the 35\,day cycles are generally 
in synchronization. This synchronization did apparently break down temporarily 
during the long \textsl{Anomalous Low} (AL3) which Her X-1 experienced in 
1999/2000, in the sense that there must have been one extra \textsl{Turn-On} cycle.
Our working hypothesis is that there are two clocks in the system, 
both with a period of about $\sim$35\,days: precession of the accretion disk (the less 
stable  ``\textsl{Turn-On} clock") and free precession of the neutron star (the more stable 
``\textsl{Pulse profile} clock"). We suggest that free precession of the 
neutron star is the master clock, and that the precession of the accretion disk is
basically synchronized to that of the neutron star through a feed-back mechanism 
in the binary system. However, the \textsl{Turn-On} clock can slip against its master
when the accretion disk has a very low inclination, as is observed to be the case 
during AL3. We take the apparent correlation between the histories of the \textsl{Turn-Ons}, 
of the \textsl{Anomalous Lows} and of the pulse period evolution, with a 5\,yr quasi-periodicity, 
as evidence for strong physical interaction and feed-back between the major 
components in the system. We speculate that the 5\,yr (10\,yr) period is 
either due to a corresponding  \textit{activity cycle} of HZ~Her or a  \textit{natural
ringing} period of the physical system of coupled components. The question whether 
free precession really exists in neutron stars is of great importance for the understanding
of matter with supra-nuclear density.}

\keywords{stars: binaries:general, --
                 accretion, accretion disks, --
                 stars: Her~X-1, --
                X-rays: general,  --
                X-rays: X-ray binary pulsars, --
                precession
               }

   \maketitle

\section{Introduction}

The binary X-ray pulsar Her~X-1 shows a number of periodic modulations of its
X-ray flux: the 1.24\,s pulse period, the 1.70\,d orbital period (through eclipses and 
the Doppler modulation of the pulse period), a 1.62\,d dip period, and a 35\,d 
super-orbital period. The latter period is observed first through an on-off cycle with a 
10\,d \textsl{Main-On} and a 5\,d \textsl{Short-On}, separated by two 10\,d 
\textsl{Off}-states (\cite{tananbaum:1972}), and second through a reproduced change 
of the shape of the 1.24\,s pulse profile (\cite{truemper:1986, deeter:1998, scott:2000}). With 
respect to these modulations we will argue that there are two $\sim$35\,d clocks in the 
system which are generally synchronized, but which were observed to temporarily 
loose synchronization during the long \textsl{Anomalous Low} in 1999/2000. 
\textsl{Anomalous Lows} seem to appear quasi-periodically about every 5\,yrs
(\cite{staubert:2006}), they are believed to be connected with episodes of low
tilt of the accretion disk, which is then blocking the line of sight to the X-ray source.

The 35\,day modulation of the X-ray flux is generally explained by the precession of 
the accretion disk which regularly blocks the line of sight to the X-ray emitting regions 
near the magnetic poles of the neutron star (\cite{gb:1976, schandl:1994}). With 
regard to the systematic variation of the shape of the X-ray pulse profiles we follow 
Tr\"umper et al. (1986), Shakura et al. (1998) and Ketsaris et al. (2000) in assuming 
that the responsible physical mechanism is free precession of the neutron star.

Free precession may appear as a fundamental physical property of rigid 
non-spherical spinning bodies. The simplest case is a spheroid with 
some small oblateness (a ``two-axial" body) in which the axis of angular velocity 
is not aligned with any principle axis of inertia (e.g., \cite{klein:1910}). 
It has been suggested as the underlying  reason
for the long-period variations, both in timing and spectral properties, 
observed in several neutron stars  (\cite{jones:2001, cutler:2003, link_epstein:2001, 
haberl:2006}). The candidate objects are mostly radio pulsars (including the Crab 
and Vela pulsars), the isolated X-ray pulsar $RX J0720.4-3125$
(\cite{haberl:2006}) and the accreting binary X-ray pulsar Her~X-1.
The existence of free precession in  neutron stars and its consequences for
our understanding of the physics of the interior of neutron stars is extensively 
discussed in the literature (\cite{anderson_itoh:1975,shaham:1977,
alpar_oegelman:1987,sedrakian:1999,wasserman:2003,levin_dangelo:2004,
link:2007}). Recently, Link (2007) has emphasized that the question of the
reality of free precession in neutron stars has strong implication for our
understanding of the properties of matter at supra-nuclear densities.

% Fig. 1
\begin{figure}
% \resizebox{\hsize}{!}{\includegraphics{PP_vs_phase.eps}}
 \resizebox{1.0\hsize}{!}{\includegraphics{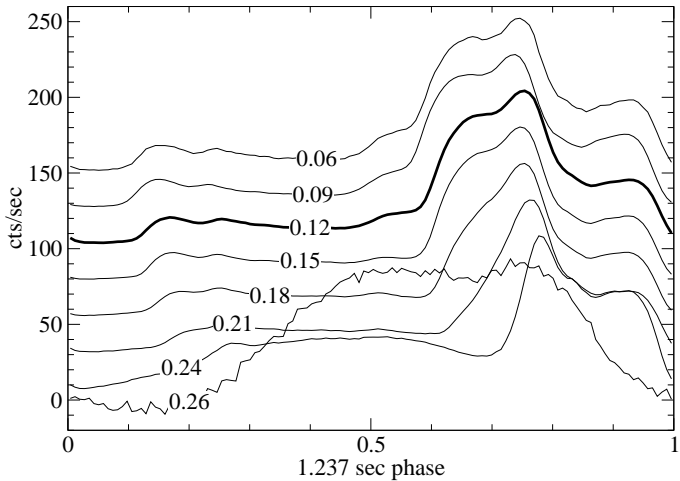}}
 \caption{Pulse profiles of Her~X-1 as observed by RXTE/PCA (3-20\,keV) during the 
 Main-On of November 2002 (35\,d cycle no. 323$^{1}$, according to the pulse 
 profile counting) as a function of 35\,d phase. For better visibility the profiles are 
 scaled to a common amplitude and shifted against each other according to their 
 35\,d phase . The profiles were generated using 128 phase bins, the curves are
 the straight line connection between adjacent data points (which are not shown).}
 \label{fig:pp_phase}
\end{figure}

% Fig. 2
\begin{figure}
%\resizebox{\hsize}{!}{\includegraphics[angle=-90]{PP_Chi_squa.eps}}
\resizebox{1.0\hsize}{!}{\includegraphics{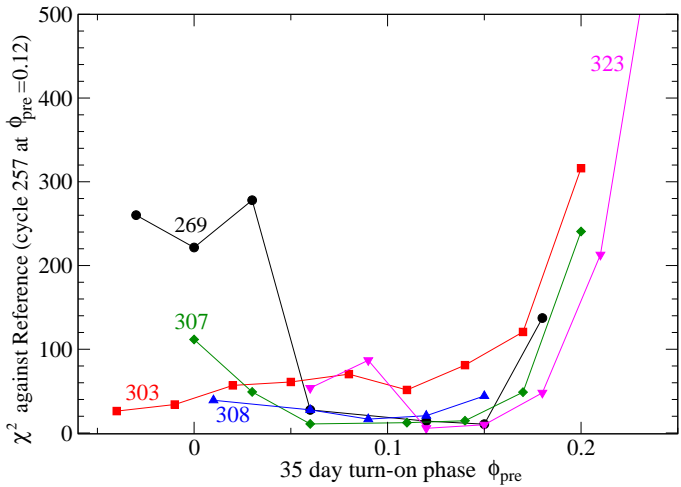}}
\caption{Comparison of pulse profiles observed in a number of 35\,day
cycles at different 35\,day phases to a template profile of cycle no. 257 at 
phase 0.12. The matching to the template profile is measured by the respective
$\chi^{2}$-value.}
\label{fig:pp_chi}
\end{figure}

In Her~X-1 the case for free precession is as long standing 
(\cite{brecher:1972, truemper:1986, shakura:1998}), as it is criticized on 
various grounds (e.g., \cite{bisnovatyi:1989}). Here we provide new observational
clues for a very stable clock which governs the regularly re-appearing pulse profiles.
We identify the stable clock with free precession of the neutron star and suggest
that the accretion disk with its rather unstable \textsl{Turn-On} clock is slaved to
the neutron star on long time scales through a closed loop physical feedback in 
the binary system.

\section{Pulse profile cycle counting}
The shape of the pulse profiles of the 1.24\,s X-ray pulsations is known to vary in 
several different ways, e.g., as a function of energy (\cite{gruber:1980}), and 
as a function of 35\,d phase (\cite{truemper:1986, soong:1990, deeter:1998, scott:2000}).
Here, we are concerned with the latter, the systematic variation with  35\,d phase. 
In Fig. \ref{fig:pp_phase} we show the result of our analysis for the Main-On 
state of November 2002 observed by \textsl{RXTE}. Our working hypothesis 
is that this modulation is due to a precessing neutron star:
our viewing angle towards the X-ray emitting polar cap region of the neutron
star varies with the phase of the neutron star precession.

%Fig. 3
\begin{figure}
% \resizebox{\hsize}{!}{\includegraphics{fig3.eps}}
\resizebox{1.03\hsize}{!}{\includegraphics{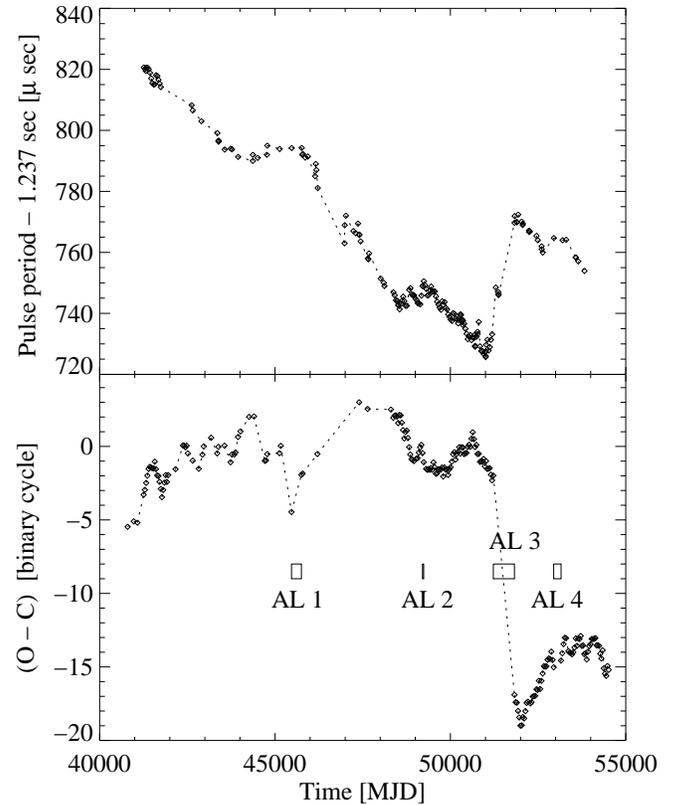}}
\caption{Pulse period and \textsl{Turn-On} histories of Her X-1 (an update of Fig.~1 of \cite{staubert:2006}).}
\label{fig:PP_O-C}
\end{figure}
\vspace*{-2mm}
\footnote{the counting of 35\,d cycles follows the convention used by 
\cite{staubert:1983}: $(O - C)$ = 0 for cycle no. 31 with turn-on near
JD 2442410}

Using observations by \textsl{RXTE} from 1996 until 2005 we have verified
that the shape of the pulse profiles is reproduced every $\sim35$ days. 
A careful timing analysis was performed of all archived \textsl{RXTE} 
data on Her X-1 
and pulse profiles were generated by folding with the measured pulse periods.  
Fig. \ref{fig:pp_phase} shows a set 
of pulse profiles (PCA, $3-20$\,keV) from 35\,d cycle no. 323$^{1}$ 
(Nov 2002) for eight different 35\,d phases. The variation of the pulse shape 
is evident. Using a profile from cycle no. 257 at 35\,d phase 0.12 as a
template (quite similar to the phase 0.12 profile of cycle no. 323, highlighted in
Fig. \ref{fig:pp_chi}), we have performed a comparison to profiles from other 
cycles. For this purpose all profiles were normalized using the amplitude of the 
main peak. Then the profiles were aligned in pulse phase: within one 35\,d
main-on this is assured by phase connection, the alignment of profiles from 
different 35\,d cycles was done using the ``sharp edge" at the decay of the 
shoulder to the right of the main peak, which was found to be the sharpest 
and most stable feature of the Her~X-1 pulse profile.
Then the difference between the count rates in the 128 phase bins were taken, 
squared and summed.  This $\chi^{2}$-sum is taken as a quantitative measure 
for the `matching' of the individual profiles to the template.
There are profiles from 16 different 35\,d cycles taken by \textsl{RXTE} over 8.8 years.
However, only in 7 cycles pulse profiles for 4 or more different 35\,d phases are available. 
In Fig. \ref{fig:pp_chi} the $\chi^{2}$-values are plotted for those 5 cycles for which 
profiles at 7 or more different phases are available. 
Each data point corresponds to a pulse profile generated from data of one 
complete day of \textsl{RXTE} observations. The center time of the observing 
interval is translated into a 35\,d phase (using a constant period of 34.85 days).
The minimum in $\chi^{2}$ for the different cycles is generally found 
around 35\,d phase 0.12
(the phase of the template), demonstrating that the profiles are repeating 
regularly. The $\chi^{2}$ comparison was repeated with three other templates 
(from different 35\,d cycles and different phases), yielding the same result.
This establishes a method of Cycle Counting which is solely based on the 
shape of the pulse profile. 

%Fig. 4
\begin{figure}
 \resizebox{0.9\hsize}{!}{\includegraphics{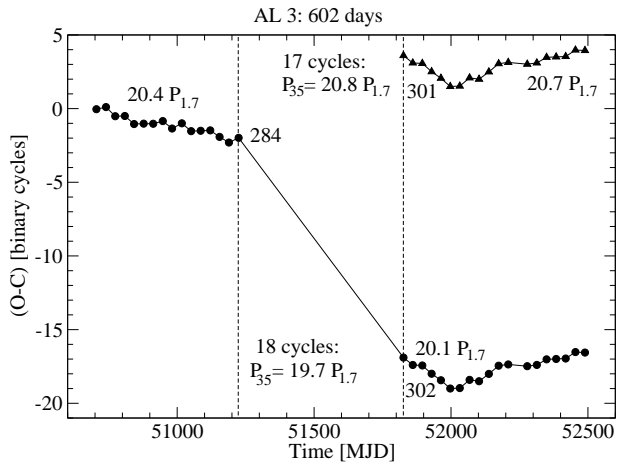}}
 \caption{The \textsl{Anomalous Low} AL3 must be bridged by 18 (not 17)
 \textsl{Turn-On} cycles: from cycle no. 284 to cycle no. 302.}
 \label{fig:AL3}
\end{figure}

\section{\textsl{Turn-On} cycle counting} 

The second method of \textsl{Cycle Counting} uses the well established
35\,d modulation of the X-ray flux (\cite{giacconi:1973}): the X-ray flux
increases sharply at \textsl{Turn-On}, reaching a maximum of the 
\textsl{Main-On} which lasts for a few days, before fading slowly into the first 
minimum. After that a second, substantially lower maximum, the 
\textsl{Short-On}, emerges for a few days, after which a second minimum
concludes the cycle (see \cite{klochkov:2006}). It is generally assumed that 
the modulation is due to shading by the \textsl{precessing accretion disk} 
(\cite{gb:1976}). There is a \textsl{Turn-On} roughly every 35 days. However, 
this clock is not very accurate: the length of a particular cycle may be longer 
or shorter than the previous one by $\sim$0.85\,days ($= P_\mathrm{orb}/2$). 

The irregularity of the \textsl{Turn-On} clock is demonstrated by the $(O - C)$ 
diagram (Fig. \ref{fig:PP_O-C}, lower panel). Here the difference between the 
\textsl{observed} turn-on time (O) and the \textsl{calculated} turn-on time (C) is 
plotted against time (\cite{staubert:1983, staubert:2000, still:2004, staubert:2006}). 
To calculate the turn-on time, a constant period of 34.85\,d is 
used (equal to $20.5 \times P_\mathrm{orb}$, with  $P_\mathrm{orb}$ = 1.700\,d). 
Fig. \ref{fig:PP_O-C} is our latest update of this diagram. If $(O - C)$ is measured 
in units of $P_\mathrm{orb}$, all data points fall more or less on horizontal lines 
(spaced by $0.5\times P_\mathrm{orb}$), due to the observed fact that the 
\textsl{Turn-Ons} occur close 
to binary phases 0.25 or 0.75. \cite{staubert:1983} (at a time when only data just 
beyond the first \textsl{Anomalous Low} - AL1- were available) had postulated that 
the change in $(O - C)$ from one cycle to the next should be either 0 or +1 or -1 in 
units of $P_\mathrm{orb}$/2 (corresponding to a cycle length of 20.5 or 21 or 20 binary 
periods). Even though the short-term development was successfully modeled by 
a random walk process, they argued for the possibility that the global long-term 
development of the diagram might be nearly flat, indicating some sort of a 
``back-driving force" which would prevent the wandering off to one or the other 
side. These assumptions have been found to hold until the occurrence of the 
dramatic event of the \textsl{Anomalous Low of} 1999/2000 (AL3). 

Fig. \ref{fig:PP_O-C} demonstrates that the \textsl{Turn-On} clock is
quite noisy, with additional quasi-periodic variations on a 5 year
time scale. $(O - C)$ correlates with the appearance of the \textsl{Anomalous
Lows} (AL), and it also strongly correlates with the neutron star's spin period 
(\cite{staubert:2006}, 2008b).

\section{Difference between pulse profile counting and \textsl{Turn-On} counting}

The two ways of cycle counting are normally consistent with one another.
However, during the long \textsl{Anomalous Low} in 1999/2000 (AL3 in 
Fig.\ref{fig:PP_O-C} ) the synchronization was apparently lost.

\begin{figure}
% \resizebox{\hsize}{!}{\includegraphics{PP_counting.eps}}
 \resizebox{\hsize}{!}{\includegraphics{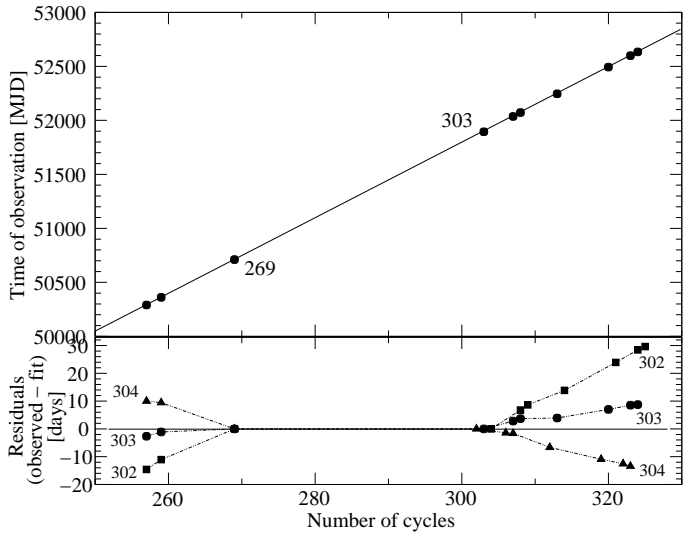}}
 \caption{Cycle counting by pulse profile fitting. Upper panel: absolute times
 of  $\chi^{2}$-minima obtained through template fitting versus cycle number,
(assuming that the gap is between cycle nos. 269 and 303). 
 Lower panel: residuals against the straight line connection of cycles 269 and 
 303 (center curve) or for the two cases where the first cycle observed after
 AL3 is called 302 or 304, respectively. The uncertainties in time are generally
 less than 0.5\,d (smaller than the size of the symbols).}
\label{fig:pp_counting}
\end{figure}

In Fig. \ref{fig:pp_counting} we plot the absolute times of the $\chi^{2}$-minima
(see Figs. \ref{fig:pp_phase} and \ref{fig:pp_chi}) found when the pulse profiles 
observed in the respective \textsl{Main-Ons} were compared to the template profile 
of cycle no. 257, against the cycle number (from pulse profile counting). The last
\textsl{Main-On} before AL3 in which good pulse profiles were obtained belongs to 
cycle no. 269, the first \textsl{Main-On} after AL3 is cycle no. 303.
The straight line in Fig. \ref{fig:pp_counting} (upper panel) is the connection
between the data points of cycle no. 269 and cycle no. 303. Dividing the difference
of the corresponding absolute times for the  $\chi^{2}$-minima by 34 (= 303 - 269)
leads to a cycle length of $34.95 \pm 0.01$\,d. The lower panel in 
Fig. \ref{fig:pp_counting} shows the residuals of the data points to the straight
line: the center curve is for the proposed cycle counting (269/303). Any different
counting to bridge the gap (e.g., 269/302 or 269/304) can be ruled out.
So, Fig. \ref{fig:pp_counting} establishes that the reference pulse shape (our 
template at 35\,d phase 0.12) does regularly repeat and that the corresponding 
observing times can be associated with unique cycle numbers. 
A linear fit to the data point in the upper pannel of Fig. \ref{fig:pp_counting}
yields a mean period of $34.98 \pm 0.01$\,d (= $20.58\cdot P_\mathrm{1.7}$)
from pulse profile counting (repeating the exercise with three other template
profiles yields the same result within uncertainties).

%Fig. 5

%\vspace*{-3mm}
Fig. \ref{fig:AL3} shows the $(O - C)$ diagram around AL3. \textsl{Turn-Ons} 
are not observed during the AL, since the source is strongly obscured 
for a total of 602 days. Assuming that the physics behind this counting, that is 
the precession of the accretion disk, is continuing during the AL, one can 
estimate the number of cycles during AL3. There are two possible solutions: 
17 cycles with a mean period $20.8\cdot P_\mathrm{1.7}$ or 18 cycles with 
a mean period of $19.7\cdot P_\mathrm{1.7}$. We conclude that there 
\textsl{must have been 18 cycles}, not 17, because of the following arguments: 
\begin{enumerate}
\item Before AL3 the mean period was low 
($20.4\cdot P_\mathrm{1.7}$), as it always is the case when going into an AL 
(\cite{coburn:2002}). Coming out of AL3 the mean period of the next 6 cycles is 
even lower ($20.1\cdot P_\mathrm{1.7}$). It is unreasonable to assume 
that the period was larger in between. 
\item The strong long-term anti-correlation between $(O - C)$ and the neutron 
star's spin and the observed dramatic spin-down during AL3 requires that $(O -C)$ 
needs to drop during AL3 (\cite{staubert:2006}), equivalent to a short period. 
\item It is believed that the \textsl{Anomalous Low} is caused by the blocking of 
the line of sight to the X-ray emitting region by the accretion disk due to a 
decreased tilt of the disk with respect to the orbital plane.
\end{enumerate}

The decrease in tilt angle is likely due to a reduced mass transfer rate,
leading to a reduced momentum of the stream onto the disk, which is
equivalent to a reduced breaking of the precession of the disk and
consequently a higher precession frequency (\cite{klochkov:2006}). 

\textsl{We conclude that during AL3 the accretion disk did one extra cycle}
in comparison to the regular cycle counting using the pulse profiles.

\section{Summary of observational facts}

Before entering the discussion we summarize the observational facts
relevant to our conclusion about the reality of free precession of the
neutron star in Her~X-1.

\begin{enumerate}
\item The pulse profile shape varies with 35\,d phase and repeats regularly.
The observations by \textsl{RXTE} over $\sim$9\,yrs establish a stable
clock with a period of $(34.98 \pm 0.01)$\,d.

\item The \textsl{Turn-On} clock is rather unstable: it shows a quasi-periodic
variation in $(O - C)$ with a $\sim$5\,yr period and an amplitude of
$\sim \pm 3$$P_\mathrm{orb}$, with `substructure' and  additional noise, and
a large step in $(O - C)$ during the 1999/2000 \textsl{Anomalous Low}
(AL3), in correlation with a dramatic spin-down (Fig.\ref{fig:PP_O-C}).

\item \textsl{Anomalous Lows} (ALs) appear quasi-periodically every
$\sim$5\,yrs, in correlation with minima in $(O - C)$.

\item $(O - C)$, the \textsl{Turn-On} history, is strongly correlated to the 
pulse period evolution (Fig.\ref{fig:PP_O-C}, and \cite{staubert:2006}).

\item The counting of 35\,d cycles by \textsl{Turn-Ons}, the $(O - C)$ diagram,
is generally synchronized (with deviations of up to $\sim \pm 3$$P_\mathrm{orb}$)
to the counting of 35\,d cycles using pulse profiles.

\item During the long AL of 1999/2000 (AL3) the \textsl{Turn-On} period was so
low, that one extra cycle was done compared to the pulse profile counting.

\end{enumerate}

\section{Discussion and conclusions}

We conclude that there are two clocks in Her X-1, both with a period of 
about 35\,d: \textit{precession of the accretion disk} and 
\textit{free precession of the neutron star}. The precessing
outer rim of the accretion disk regularly blocks the line of sight to the X-ray
emitting polar caps of the neutron star, thereby producing the \textsl{Turn-On}
cycle. The free precession of the neutron star is responsible for the orientation
of the beamed X-ray emission, thereby producing the periodic modulation
of the shape of the observed pulse profiles. 

The two clocks are so similar 
in period because of synchronization due to strong feed
back in the system (it would seem unreasonable to assume that the two
clocks have nearly the same frequency purely by chance).
We propose that free precession of the neutron star is the 
\textsl{master clock} which is rather stable (it could however change rapidly due 
to a change in the shape of the neutron star, e.g., as a result of a star quake).
The accretion disk precession is observed to be a noisy clock with some
systematic variations.
There are quite a number of torques acting on the accretion disk: the tidal 
force from the optical star, the internal viscous force, dynamical forces
due to the impact of the accretion stream and the illumination by the X-ray beam, 
and, finally, the neutron star magnetosphere interacting with the inner edge 
of the disk. These forces collectively produce the precession (as well as the tilt
and the warp) of the disk (\cite{shakura:1998, klochkov:2006}). We consider the 
tidal force and the dynamical forces to be the dominating ones. In the absence 
of dynamical forces the precession would be much faster, with a period around 
15\,d (\cite{shakura:1999}). 

So the question is: how does the neutron star with its inherent period of free 
precession of 35\,d manage to ``slave" the accretion
disk? We believe that the critical parameter for the closed loop feedback
in the system is the rate of mass transfer from the optical star. The surface 
of HZ~Her facing the neutron star is illuminated and heated by the neutron 
star's X-ray emission, enhancing the mass transfer. However, the heating
is not constant and uniform because the accretion disk blocks part of the
X-ray beam and modulates (spatially and temporally) the heating of the optical
star's surface according to its precessional motion. A first loop may be the 
following: the inner part of the accretion disk follows the free precession of
the neutron star (because of the magnetospheric interaction), the shadowing
of the optical star from the X-ray beam follows the movement of the inner disk,
modulating the mass transfer rate with the period of the neutron star free 
precession. The accretion stream hits the accretion disk applying a force to
slow down the disk's precession and to influence its tilt, as shown by 
numerical simulations (\cite{shakura:1999,klochkov:2006}). A variable tilt of the
outer disk results again into a variable shading of the optical star. Finally,
the variable mass transfer rate will eventually (after a viscous time scale) 
show up in a variable mass accretion rate which governs the X-ray luminosity.
The X-ray luminosity, on the other hand, is crucial for the heating of the
optical star as well as for the illumination of the outer parts of the accretion disk
where coronal winds and torques may be produced (\cite{schandl:1994}).
With the variable heating of the optical star by the X-ray beam, both through
variable shading and through variable X-ray flux, the loop is closed.

Note also, that the strong correlation between the $(O - C)$ diagram and
the pulse period evolution (Fig.\ref{fig:PP_O-C} and \cite{staubert:2006}) 
can be understood within this model: any enhanced mass accretion rate
(because of an enhanced mass transfer rate) will result in an enhanced
angular momentum transfer, and hence a spin-up.

We propose that the described physical couplings provide 
strong feed-back mechanisms in the Her~X-1/HZ~Her binary system which 
establish a delicate equilibrium of the whole system.
We assume that the physical coupling described above is strong enough
to lock the precession of the outer accretion disk to that of the neutron star 
(this may not be possible if the natural frequencies of the precession 
of the neutron star and that of the accretion disk were different by much more
than the estimated factor of about 2). Due to the large number of forces acting on 
the accretion disk (probably all of them being subject to noise) one may 
understand that the synchronization between the two clocks is not perfect: 
First, there is the modulation of the \textsl{Turn-On} times. We know now, that 
random walk is not the right model for this modulation. There seems to be just 
random noise super-imposed onto the quasi-periodic up and down in $(O - C)$. 
Over long times the deviations are limited to $\sim \pm 3$$P_\mathrm{orb}$. In our 
current model we would now interpret the ``back-driving force", postulated by 
Staubert et al. (1983), as being due to the coupling of the accretion disk to
the neutron star. Second, the dramatic event observed during AL3, in which
the accretion disk showed a low tilt and a fast precession, could then
be viewed as an extreme behavior, demonstrating that the accretion
disk has a ``life of its own", and it is able to temporarily escape the slaving 
by the neutron star. It seems, however, that the equilibrium is 
re-established rather quickly. Note that in quoting a 35\,d
cycle no. for the time after AL3 one has to clearly state what method of cycle 
counting it refers to: \textsl{pulse profile counting} or \textsl{Turn-On counting}, 
the latter is advanced by one cycle. 

We also like to draw attention to the fact that the $(O - C)$ diagram 
(Figs. \ref{fig:PP_O-C} and  \ref{fig:AL3}) has - averaged over time 
scales $>15$\,yrs - a positive slope from the time of discovery of the 
source until today, if \textsl{pulse profile counting} is used.
In this case the diagram continues after AL3 with the upper 
right curve in Fig. \ref{fig:AL3}, corresponding to the solution with 17
neutron star cycles inside AL3. A linear fit to all data of the $(O - C)$ 
diagram in \textsl{pulse profile counting} yields a mean cycle duration of 
($34.8791\pm0.0001$)\,d, which we associate with the long-term average
period of the neutron star precession. Using only data after AL2, we find
an average period of ($34.9681\pm0.0003$)\,d, quite close to the period 
of ($34.98\pm0.01$)\,d, found from pulse profile fitting (over a similar 
observational period). The quoted uncertainties, however, are statistical 
uncertainties only, for the true physical uncertainties one would have to 
add systematic uncertainties due to the irregular modulation over the 
limited time base and the non-uniform sampling of $(O - C)$. So, we 
refrain from speculating about variation of the period over time.
We take the above finding as independent support for the already reached 
conclusion that the precession of the accretion disk follows that of the 
neutron star on long time scales. We predict, that the mean upward trend 
in $(O - C)$ will continue in the future. 

The key feature of our model is that free precession of the neutron star 
is responsible for the observed long-term stability of the 35\,d cycle
(both, the regular re-appearance of pulse profiles as well as the long-term
turn-on history). Unlike in isolated neutron stars, where free precession is 
damped by dissipative processes, the free precession of the old neutron 
star in Her~X-1 can be sustained for long times by the accretion feedback 
loop described above, which may lead to quite different properties.
We also note that Lamb et al. (1975) have already concluded that
phase dependent torques are capable of exciting (or damping) large
amplitude neutron star wobble.

With regard to the apparent quasi-periodic 5\,yr  (10\,yr) modulation in 
$(O - C)$, seen in correlation with the pulse period evolution and the 
appearance of the \textsl{Anomalows Lows}, we have no definite answer.
We see two possibilities (\cite{staubert:2006}).: either, the
modulation is due to an ``activity cycle" of HZ~Her changing the mass
transfer rate (see also \cite{still:2006}), alternatively, the $\sim$5\,yr
may represent a natural ringing frequency of a system of several coupled 
physical components. 

We like to address a final question: are shifts in pulse arrival time
observed which are expected to occur for a precessing pulsar?
A decomposition analysis of high quality pulse profiles observed 
with \textsl{RXTE} using eight Gaussian components shows that the main 
peak (as well as other peaks) is (are) systematically varying in amplitude and in 
relative phase (e.g., with respect to the well defined minimum). A quantitative
description is in preparation. In addition, we are making progress in
modeling the changing shape of the pulse profiles. First results, assuming a 
spot-like emission region at each of the two magnetic poles and additional arc-like 
emission structures around the poles, were published by Wilms et al. (2003) 
and again reported by Postnov (2004). In the case of a freely precessing
neutron star radiating like a pulsar, systematic variations in the observed
period and pulse phase with 35\,d phase are expected 
(\cite{ruderman:1970, shakura:1988, postnov:1991, bisnovatyi:1993}), 
with a maximum relative change in period
of the order $10^{-6}$. Any change in pulse period will result in a shift of the
pulse arrival time. Unfortunately such shifts are observationally 
indistinguishable from shifts resulting from period variations 
due to accretion torque changes. The pulse period history shown in 
Fig.~\ref{fig:PP_O-C} (upper panel) shows strong and frequent period variations on 
time scales ranging from 18\,d  to beyond 1000\,d. The strongest variations reach
 dP/dt-values of nearly $\pm3\times10^{-12} ss^{-1}$ (observed at the smallest
 time scales), corresponding to a relative change of $10^{-6}$ over a few days. 
 In addition, detailed pulse arrival time analysis of \textsl{RXTE} data has shown 
 that similar changes (of both signs) are found on time scales of a few days
 (see also \cite{klochkov:2008}). 
 Our six measured values  do not show any correlation with
 35\,d phase.  We attribute these to changes in mass accretion rate.

In summary, we conclude that our analysis does support the idea of
free precession to be present in the neutron star of Her~X-1. Our main line
of arguments rest on the identification of two different  35\,d clocks in this 
system: free precession of the neutron star (as the master clock) and precession 
of the accretion disk (which is quasi-synchronized to the neutron star for most 
of the time). Long-term shifts in pulse arrival time, as seen in radio pulsars, are 
principally non-observable because of the always present accretion torque noise 
and resulting pulse frequency variations. However, the observed systematic 
relative shifts of structures in the pulse profile with corresponding changes in 
pulse width and amplitude as a function of precessional phase are reminiscent 
of precessing radio pulsars. Link (2007) has concluded that the standard picture
of an outer core of a neutron star consisting of coexisting superfluid neutrons
and type II superconducting protons is inconsistent with the existence of long-period
precession. Free precession then means that the neutron vortices at the inner 
crust must not be pinned, in agreement with similar conclusions reached by 
Shaham (1977) and Jones \& Anderson (2001).

\begin{acknowledgements}
We acknowledge the support through DFG grant St 173/31
and the corresponding RFBR grant 06-02-16025 as well as significant
contributions to the data analysis by L. Rodina. We thank the
anonymous referee for pointing to the question of shifts in pulse arrival
time.  We are thankful to J. Tr\"umper for suggesting to emphasize the 
general importance of free precession to the physics of the neutron star 
interior.
\end{acknowledgements}

%\bibliographystyle{aa}
%\bibliography{refs}

\end{document}